\documentclass[usenatbib]{mn2e}
\usepackage{graphicx}                                            
\usepackage{times}
              
\title[K-correction for GX 339-4]{On the masses and evolutionary status of the black hole binary\\
 GX 339-4. A twin system of XTE J1550-564?}

\author[Mu\~noz-Darias, Casares \& Mart\'\i{}nez-Pais]
{T. Mu\~noz-Darias$^1$ \thanks{E-mail: tmd@iac.es},
J. Casares$^{1}$ and I.G. Mart\'\i{}nez-Pais$^{1,2}$\\
$^1$ Instituto de Astrof\'{\i}sica de Canarias, 38200 La Laguna, Tenerife, Spain\\
$^2$ Departamento de Astrof\'{\i}sica, Univ. de La Laguna, E-38206 La Laguna, Tenerife, Spain\\}
\begin{document}

\maketitle

\begin{abstract}
We apply the K-correction to the black hole LMXB GX 339-4 which implies $M_{X}\geq 6 M_{\odot}$ by only assuming that the companion is more massive than $\sim 0.17M_{\odot}$, the lower limit allowed by applying a 'stripped-giant' model. This evolutionary model successfully reproduces the observed properties of the system. We obtain a maximum mass for the companion of $M_2 \leq 1.1 M_{\odot} $ and an upper limit to the mass ratio of $q(=M_2/M_{X})\leq0.125$. The high X-ray activity displayed by the source suggests a relatively large mass transfer rate which, according to the model, results in $M_2 \ga 0.3M_{\odot}$ and $M_{X} \ga 7M_{\odot}$. We have also applied this scenario to the black hole binary XTE J1550-564, which has a similar orbital period but the donor is detected spectroscopically. The model successfully reproduces the observed stellar parameters.
\end{abstract}


\section{Introduction} \label{introduction}
Low Mass X-ray Binaries (LMXB) are interacting binaries harbouring a neutron star (NS) or a black-hole (BH) which accretes matter from a low mass star. 
A subgroup, the so-called X-ray transients (SXTs), provide a unique opportunity to set dynamical constraints to the the masses of both NSs and BHs because the spectrum of the companion dominates during the quiescent (X-ray off) states. In particular, we can empirically determine the mass function of the system which is usually expressed as:
\begin{equation}
\label{mf}
f(M)=\frac{PK_2^3}{2 \pi G}=\frac{M_X\sin{i}^3}{(1+q)^2}
\end{equation}
where $P$ is the orbital period, $K_2$ the semi-amplitude of the velocity curve of companion star, $M_X$ the mass of the compact object, $i$ is the orbital inclination of the systems and $q=\frac{M_2}{M_X}$  the mass ratio. Therefore, it is possible to establish a secure lower limit to the mass of the compact object by only measuring $P$ and $K_2$.
Although this method has successfully been applied to $\sim 20$ BH X-ray binaries (see \citealt{C2006} for details), they only represent the 'peak of the ice-berg' of a large population of $10^8-10^9$ stellar-mass BH present in our galaxy (e. g. \citealt{BB94}). Most quiescent SXTs are optically faint, with $R\geq 20$, and hence dynamical constraints are usually affected by large errors. Setting new dynamical constraints and refining the existing mass solutions are the only ways to get new insights into the fundamental properties of stellar-mass BHs.\\
GX 339-4 (V821 Ara) was discovered by OSO 7 in 1972 (\citealt{Ma73}). Since then, several X-ray outburst (e. g. 4 in the last decade) have been observed, and all the X-ray states have been detected in this system (See \citealt{Mvdk97} for the intermediate state and \citealt{Mi91} for the Very High state). Moreover, \cite{Cor00} detected a compact jet emission from this system, showing that it is a microquasar. GX 339-4 was early proposed as a BH candidate based on its X-ray properties (\citealt{S79}) but the spectral features of the companion star could not be detected even during 'X-ray off' states (\citealt{Tariq01}), preventing a dynamical confirmation (i. e. $M_X > 3M_{\odot}$). Only during the 2002 outburst \cite{hynes03} reported the first detection of the donor star thanks to the discovery of NIII/CIII Bowen emission lines arising from the irradiated companion star. The lines are very sharp and swing with a velocity semi-amplitude of $K_{em}$=$317 \pm 10 $ km s$^{-1}$. These authors also reported an orbital period of $P=1.7557\pm0.0004$ days which was later confirmed by \cite{lyc06}. 
The combination of both, radial velocity of the companion and orbital period yields $f(M)=5.8 \pm 0.5 M_{\odot}$ (and hence $M_X \geq 5.3M_{\odot}$), which represents the first dynamical proof for a BH in GX 339-4. Note that narrow high-excitation emission lines originating from the companion star have also been detected in many others LMXBs (e.g. \citealt{C04})
and demonstrate that this is a common feature in LMXBs with high X-ray activity (i. e. steady systems and transients during outbursts). However, the NIII/CIII emission lines are excited on the inner hemisphere of the donor star by the Bowen mechanism/photoionization (\citealt{MCT75}) and only provides a lower limit ($K_{em}$) to the true $K_{2}$-velocity of the donor. In \cite{MCM05} we tackle this problem in a general approach by modeling the deviation between the reprocessed light-center and the center of mass of a Roche lobe filling star (the so-called 'K-correction') including screening effects by a flared accretion disc. In this paper we compute the K-correction for GX 339-4 which provides a more restrictive lower limit to the mass of the BH in this binary. In sections \ref{companion} and \ref{observations} we discuss the nature of the companion star and show that all the known observables can be explained by an scenario in which a 'stripped-giant' companion transfers mass onto a BH. We also apply this evolutionary model to XTE J1550-564 which shows observational properties very similar to GX 339-4. The stellar parameters of the companion in this system are in excellent agreement with a stripped-giant scenario.  

\section{K-correction for GX 339-4} \label{kcor}

\begin{figure}
  \includegraphics[width=8cm]{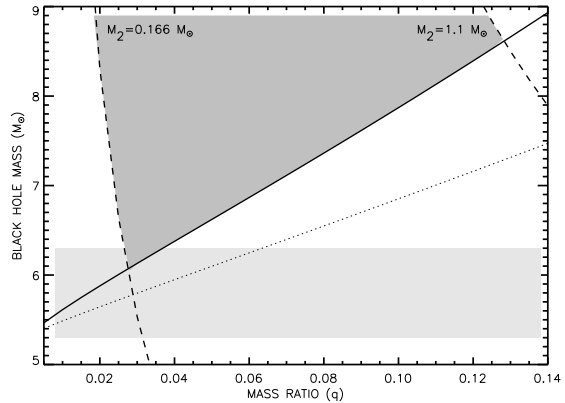}
  \caption{K-correction for GX 339-4. We show a strict lower limit to $M_{X}$ (thick solid line) as function of the mass ratio (q) and the mass of the companion($M_2$). We find $M_X \geq 6.0{M_\odot}$ (thick solid line) and $q \leq 0.125$. The dotted line shows the same limit by considering only the value reported by H03 ($M_{X}\geq 5.3 M_{\odot}$), the absence of X-ray eclipses and the dependence of $f(M)$ with $(1+q)^2$ (see eq. \ref{mf}). We also mark  $f(M)=5.8\pm0.5 M_{\odot}$ reported by H03 as the light, grey region.}
\label{all}
\end{figure}

Following MCM05 the relation between the radial velocity corresponding to the center of mass of the companion ($K_2$) and the observed emission line velocity($K_{em}$) can be expressed as:
\begin{equation}
\label{eq2}
K_{2}=\frac{K_{em}}{1-f(1+q)}
\end{equation} 
where $f$ is a dimensionless factor which represents the  displacement of the emission light-center from the center of mass of the companion in units of the orbital separation. The value of $f$ depends on how much area of the irradiated hemisphere of the donor is shadowed by the flared accretion disc. According to MCM05, $f$ is constrained between
\begin{equation}
\label{limit}
0.5+0.227 \log q>f>\sin^2{\alpha_{M}}
\end{equation}
where the left hand side of the equation represents the case when the emission line is formed in the inner Lagrange point whereas the right hand side corresponds to the case where the emission arises from the irradiated point with maximum radial velocity. This point is located at the limb of the irradiated region and holds an opening angle $\alpha_{M}$ above the plane of the binary. An analytical expression of $\alpha_M$ as function of q is provided by Paczynski's equation (1971),

\begin{equation}
\label{am}
\sin{\alpha_{M}} \cong \frac{R_{2}}{a} \cong 0.462(\frac{q}{1+q})^{1/3} .
\end{equation}
It is clear from eq. \ref{eq2} that $K_2$ (and hence $f(M)$) increases with $f$. From the lower limit to $f$ in eq. \ref{limit}, it is possible to establish a more restrictive  lower limit to $M_{X}$ than the one provided by H03 $f(M)$ (i. e. $M_X \geq 5.3M_{\odot}$). In fig. \ref{all} we show the lower limit obtained for $M_{X}$ as function of $q$ (thick solid line). 
Furthermore, GX 339-4 does not show eclipses of the X-ray source which results in the following restriction to the orbital inclination ($i$): 
\begin{center}
$i \leq 90 -\alpha_{M}(q)$. 
\end{center}
By using these constraints in eq. \ref{mf} we obtain a strict lower limit to   $M_{X}$  as function of $q$. This is shown in fig. \ref{all} as a thick solid line. The dotted line shows the lower limit to $f(M)$ reported by H03 but including also the above constraint on $i$ and the dependence of $f(M)$ with $(1+q)^2$. Hence, the difference between the solid and dotted lines gives the net effect of the K-correction in the lower limit of the BH mass. For instance, it rises the lower limit by $\sim 0.5 M_{\odot}$ for $q=0.05$ and $\sim 1 M_{\odot}$ for $q=0.1$.
Unfortunately there are no strong constraints on $q$ and hence a restriction to the companion mass are clearly needed in order to derive a lower limit to $M_X$ independently of $q$. 

\section{The nature of the companion star in GX 339-4}
\label{companion}
In terms of LMXB evolution there are three possible cases in which stable mass transfer occurs. These three evolutive paths essentially depend on what time scale (angular momentum loss or stellar evolution) brings the companion into Roche lobe contact first (e. g. \citealt{KKB96}). In this picture, LMXBs with orbital periods around $\sim$ 2 days descend from systems where the companion has evolved off the main sequence as a subgiant before angular momentum losses shrink the Roche lobe enough to allow mass overflow. 
\cite{FFW72} showed that the mean density ($\bar{\rho}$) of a Roche lobe-filling companion star is determined solely by the binary period $P$:
\begin{center}
$\bar{\rho} \cong 113P_{hr}^{-2}$ g cm$^{-3}$
\end{center}
where $P_{hr}$ is the orbital period in hours.
For GX 339-4, with $P_{hr}=42.14$ hours, we obtain $\bar{\rho} \cong 0.06$ g cm$^{-3}$ which according to \cite{COX} would correspond to a B main sequence star with $\sim 17M_{\odot}$. This scenario does not fit with the GX 339-4 case since a B star would dominate the optical spectrum of the binary. On the other hand, late-type giants have even lower mean densities than the companion in GX 339-4. Therefore, the companion star is probably a subgiant, as it was previously suggested by \cite{hynes04}.\\ 

A subgiant is a low mass star which has evolved off the main sequence and posses a helium core which represents an approximate fraction $\sim 0.15$ of the total stellar mass (see \citealt{K88} for details). \cite{T83} and \cite{WRS83} have shown that all the stellar properties depend on the core mass ($M_{c}$) instead of its total mass ($M_2$). In this scenario mass transfer is driven by the expansion of the Hydrogen burning shell as consequence of the nuclear evolution. This so-called 'stripped-giant'
evolutionary model, which is valid for $M_c \la 0.45 M_{\odot}$, has been applied by \cite{K93} to the case of the BH LMXB V404 Cyg with convincing results. Using the orbital period of GX 339-4 ($P=1.7557$ d) into the K93's equations we obtain
\begin{equation}
\label{mc}
(m_c/0.25)^{7.65}m_2^{-0.5}=0.107
\end{equation}
with $m_c=M_c/M_{\odot}$ and $m_2=M_2/M_{\odot}$. According to K93 $M_c$ is constrained between 
\begin{center}
$M_2\geq M_c \geq 0.17M_2$
\end{center}
where the right hand side condition comes from the Schonberg-Chandrasekhar limit.

\begin{table}
\begin{center}
\caption{Stripped giant model for GX339-4}
\label{gx339}
\begin{tabular}{c c c }
\hline
 & Minimum Mass & Maximum Mass\\
& solution & solution\\
\hline
Core Mass $(M_{\odot})$ & 0.166 & 0.187\\
Total Mass $(M_{\odot})$ & 0.166 & 1.1\\
Radius $(R_{\odot})$ & 1.56 & 2.93 \\
Luminosity $(L_{\odot})$& 1.19 & 3.25 \\
$T_{eff} (K)$ & 4837 &  4533 \\
$-\dot{M_2} (M_{\odot}$ yr$^{-1}$) & $4.9\times10^{-11}$ & $7.8\times10^{-10}$\\
$-\dot{M_2}/\dot{M_1}$ & $0.32-2.0\times10^{-3}$ & $0.51-3.2\times10^{-2}$\\
BH mass $(M_{\odot})$ & $\geq 6$ & $\geq 8.6 $ \\
\hline
\end{tabular}
\end{center}
\end{table}

\subsection{Minimum and maximum mass for the companion}
If we use the condition $m_c=m_2$ in eq.\ref{mc} we obtain $M_c=M_2=0.166M_{\odot}$. This mass is the minimum mass permitted for the companion and represents the case in which all the Hydrogen shell is burned and the companion becomes a Helium white dwarf. As we show in fig.\ref{all} as the dashed line, this limit case constrains the mass of the BH in GX339 to $M_X \geq 6 M_{\odot}$, although this lower limit could be as high as $M_X \geq 7.2 M_{\odot}$ if we consider the H03 upper limit to $f(M)$ (i. e. $6.3 M_{\odot}$) in the K-correction computation.\\
On the other hand, we obtain $M_2\leq1.1M_{\odot}$ ($M_c\leq0.187M_{\odot}$) by applying the condition $M_c/M_2 \sim 0.17$. This is the maximum mass permitted for the companion through the stripped-giant model and sets an upper limit to the mass ratio of $q\leq0.125$ (fig. \ref{mc}). If the companion were as massive as $1.1 M_{\odot}$ the BH would be more massive than $8.6 M_{\odot}$. We list the solutions of the stripped-giant model for GX339-4 in Table \ref{gx339}.


\section{Observational constraints to GX 339-4}
\label{observations}
Following K93, the radius, luminosity and mass transfer rate of a stripped-giant Roche lobe filling companion can be calculated as function of the mass of the Helium core according to:
\begin{equation}
\label{r2}
R_2=12.55R_{\odot}(m_c/0.25)^{5.1}
\end{equation}
\begin{equation}
\label{l2}
L_2=33L_{\odot}(m_c/0.25)^{8.11}
\end{equation}
\begin{equation}
\label{m2p}
-\dot{M_2}=5.4\times10^{-9}(m_c/0.25)^{7.11}m_2
\end{equation}
For the minimum mass solution we obtain $R_2=1.56 R_{\odot}$ and $L_2=1.19L_{\odot}$ which gives a companion effective temperature $T_{eff}=4837$ K. On the other hand, if we consider the maximum mass solution we obtain $R_2=2.93 R_{\odot}$, $L_2=3.25 L_{\odot}$ and $T_{eff}=4533$ K. 
Although spectroscopic features of the companion star have not been detected yet, SFC01 established a lower limit to the r-band Gunn magnitude of the companion of $r\geq20.4$. We have computed the expected $r$ magnitudes for our derived companions as function of the distance ($d$) to GX 339-4 which is estimated to be in the range $6 \leq d \leq 15$ kpc (H04). For the calculation we have used the mean value for the reddening of $E(B-V)\sim1.2$ reported in H04. According to \cite{SK82}, $R_0=R+2.32 E(B-V)$ where $R_0$ is the dereddened magnitude which is computed following the $r$ to $R$ transformation reported in \cite{Wi91}. For $d\sim 6$ kpc we estimated $21.8 \geq r \geq 20.6$ for the minimum and maximum mass solutions respectively whereas we derive  $23.8 \geq r \geq 22.6$ if $d\sim 15$ kpc. Therefore, it is clear that a stripped-giant companion is compatible with the non-detection of the secondary by SFC01.\\
On the other hand, eq.\ref{m2p} gives a mass transfer rate of $4.9\times10^{-11} M_{\odot}$ yr$^{-1}$ and $7.8 \times10^{-10} M_{\odot}$ yr$^{-1}$for the minimum and maximum mass solutions respectively. KKB96 showed that the critical mass transfer rate ($\dot{M_{ct}}$) for a system to be a persistent source (i. e. not transient) can be estimated by using the expression 
\begin{equation}
\label{mcrit}
\dot{M_{ct}}=5\times10^{-11}(m_1)^{2/3}(\frac{P}{3hr})^{4/3} M_{\odot} yr^{-1}
\end{equation}
where $m_1$ is the mass of the compact object in $M_{\odot}$ and $\frac{P}{3hr}$ the orbital period in units of 3 hours.
If we substitute in this equation the orbital period of GX 339-4 and assuming $M_X\sim10M_{\odot}$ we obtain $\dot{M_{ct}}\sim8\times10^{-9} M_{\odot}$ yr$^{-1}$, which shows that our values obtained through the stripped-giant model are consistent with the transient nature of the source. \\
\cite{Ho05} detected a maximum X-ray flux during the 2002/2003 outburst of $F_X=3.25\times10^{-8}$ erg s$^{-1}$ cm$^{-2}$  which results in a X-ray luminosity in the range $L_X(peak) \sim 1.4-8.7\times10^{38}$ erg s$^{-1}$ depending on the assumed distance. Applying 
\begin{equation}
\label{m1p}
L_X(peak)=\eta\dot{M_1}c^2 
\end{equation}
with $\eta\sim0.1$ (\citealt{ap}) we obtain $\dot{M_1}\sim 0.25-1.54\times10^{-7} M_{\odot}$ yr$^{-1}$. Therefore, we find a predicted outburst duty cycle for the outburst in the range $-\dot{M_2}/\dot{M_1} \sim 0.32-2.0\times10^{-3}$ for the minimum mass solution and $-\dot{M_2}/\dot{M_1} \sim 0.51-3.2\times10^{-2}$ for the maximum companion mass case. Classical transient sources (e. g. V404 Cyg) only show an outburst every few decades and hence its duty cycles are low ($\sim10^{-4}$). This is clearly not the behaviour showed by GX 339-4 which has undergone four outbursts in the last $\sim 10$ years. This suggests that $\dot{M_2}$ is close to $\dot{M_{ct}}$ for GX 339-4, and therefore a companion mass in the upper part of the proposed mass range seems to better explain the high X-ray activity of this system.

 \begin{table}
\begin{center}
\caption{Companion star in XTE J1550-564}
\label{j1550}
\begin{tabular}{c c c }
\hline
 & Orosz et al. 2002 & Stripped giant model\\
\hline
Mass $(M_{\odot})$ & 0.4-1.3 & 0.16-1.09\\
Radius $(R_{\odot})$ & 1.88-2.81 & 1.42-2.67 \\
Luminosity $(L_{\odot})$& 1.4-4 & 1 - 2.8 \\
$T_{eff} (K)$ & 4100-5100 &  4574-4880 \\
$-\dot{M_2} (M_{\odot}$ yr$^{-1}$) & $0.1 - 7\times10^{-9}$ & $0.04 - 0.7\times10^{-9}$\\
\hline
\end{tabular}
\end{center}
\end{table}

\section{XTE J1550-564, a twin system of GX 339-4?}
The LMXB XTE J1550-564 was discovered on 1998 by the All Sky Monitor (ASM) onboard the \textit{Rossi X-ray Timing Explorer} (\citealt{Sm98}). \cite{Or02} showed that this system harbours a BH with $f(M)=6.86 \pm 0.71 M_{\odot}$ and $M_{X} \sim 10M_{\odot}$. These authors also classify the companion star in the range G8IV-K4III with $4100\leq T_{eff} \leq 5100$ K and $M_2\sim 0.4 M_{\odot}$ although a tentative measurement of the rotational broadening suggests $M_2\geq0.9 M_{\odot}$.  On the other hand, the orbital period is $P=1.552$ days (O02) which is very similar to that of GX 339-4. The stripped-giant model (see above equations) for this $P$ yields $0.16\leq M_2 \leq 1.09 M_{\odot}$ with $4880 \geq T_{eff} \geq 4574$ K, which is highly consistent with the observations. In table \ref{j1550} we compare the observed properties of the companion in XTE J1550-564 with those obtained through the stripped-giant model and the parameters are in excellent agreement.\\
Since its discovery in 1998, XTE J1550-564 has undergone two more X-ray outbursts suggesting a mass transfer rate close to $\dot{M_{ct}}$. O02 estimate $\dot{M_2}=0.1-7\times10^{-9} M_{\odot}$ yr$^{-1}$ which yields $\dot{M_2}/\dot{M_{ct}}\sim 1.5\times10^{-2}-1$. 
During the 1998 outburst, \cite{Ho01} measured a maximum X-ray flux $F_X=1.47\times10^{-7}$ erg s$^{-1}$ cm$^{-2}$ which results in $L_X(peak) \sim 4.94\times10^{38}$ erg s$^{-1}$ ($d\sim 5.3$ kpc; O02) and $\dot{M_1}\sim 8.72\times10^{-8} M_{\odot}$ yr$^{-1}$ (eq \ref{m1p}). By combining the observational constraints to $\dot{M_2}$ and $\dot{M_1}$ we obtain a predicted duty cycle in the range $-\dot{M_2}/\dot{M_1} \sim 0.12-8.03\times10^{-2}$. As expected, these values are larger that those typical of classical transients (e. g. V404 Cyg) which have longer outburst recurrence times. We get to reproduce these values by considering stripped-giant companions with $M_2 \geq 0.28$, for which we predict  $-\dot{M_2}$ in agreement with the observations. We want to note that XTE J1550-564 is also interesting because is one of only four SXTs for which the Very High X-ray state has been observed (e. g. \citealt{So99}; GX 339-4 is one of the other three). Moreover \cite{Ha01} observed relativistic plasma ejections at radio wavelenghts indicating that XTE J1550-564 is a microquasar as well. 

\section{Discussion} 
We have applied the stripped-giant model to the BH LMXBs GX 339-4 and XTE J1550-564. These systems share the following properties:\\
(i) Orbital periods in the range 1.5-1.7 d.\\
(ii) Transient behaviour with frequent X-ray outburst\\
(iii) The Very High state has been observed (a total of only 4 system have shown this X-ray state)\\
(iv) Both are microquasars.\\
For the case of XTE J1550-564, where the companion star has been detected and its stellar parameters constrained, we find that the proposed evolutionary model successfully reproduces all the observables. Therefore, it seems probable that XTE J1550-564 harbours a stripped-giant companion. The case of GX339 is less straightforward since the companion star has not yet been observed. Only through the detection of the NIII/CIII Bowen lines arising from the irradiated donor  a lower limit ($K_{em}\sim 317$ km s$^{-1}$) to the velocity of the companion was established by H03. We have applied the K-correction to this $K_{em}$ velocity and derived $M_X \geq 6 M_{\odot}$ including the error bars in $f(M)$ reported by H03. This value represents a solid lower limit to the mass of the BH in this LMXB. This result comes from the assumption that the companion is a stripped-giant with the minimum possible mass. Although there is not a definitive evidence for a stripped-giant donor in this system, this model is favoured by the $\sim 1.7$ d orbital period which clearly points to an evolved companion. Moreover this evolutionary model is consistent with the transient behaviour of the source and also with the large number of X-ray outbursts displayed. The radius and luminosity predicted by the stripped-giant model also explain the non-detection of the companion  by SFC01. However, the above lower limit to $M_{X}$ is quite conservative and unrealistically small since the companion is not an Helium white dwarf because it would not fill a 1.7 d Roche lobe. Furthermore, the $\dot{M_2}$ predicted by the minimum mass solution yields an outburst duty cycle which is much too long to explain the frequent X-ray activity in GX 339-4, with 4 outbursts in the last 10 years. Assuming the ratio $\dot{M_2}/\dot{M_{ct}}$ for GX 339-4 is similar than for XTE J1550-564 ($\dot{M_2}/\dot{M_{ct}}\geq 1.5\times10^{-2}$) we estimate  $\dot{M_2} \geq 1.2\times10^{-10} M_{\odot}$ yr$^{-1}$, which results in $M_2\geq0.3 M_{\odot}$. If we combine this limit with the K-correction we obtain $M_{X}\geq6.6-7.7M_{\odot}$ considering the error bar in $f(M)$. Although this lower limit is less secure than the one obtained through the minimum mass solution it is probably more realistic and better explains the frequent X-ray activity displayed by this source. Note that the stripped-giant model is only valid for $M_c \la 0.45 M_{\odot}$ and hence if we consider heavier evolved companions (e. g. giant stars) with $M_2 \geq M_c \ga  0.45 M_{\odot}$ we would obtain higher $M_{X}$ values. However, as we explain in section \ref{companion}, this possibility is at odds with the mean density of the donor derived from the orbital period.\\ 
\section{Conclusion}
We have applied the K-correction to the BH LMXB GX 339-4. By considering the limit case where the emission line is formed at the limb of the irradiated region of the companion we derive a solid lower limit to $M_{X}\geq 6 M_{\odot}$ including the error bars in $f(M)$. Here we have only assumed $M_2\geq0.166 M_{\odot}$, the lower limit allowed by the stripped giant model. We find that the stripped-giant evolutionary model explains the non-detection of the companion by SFC01 and the X-ray behaviour of the source. In particular, we propose $M_{2}\ga0.3M_{\odot}$, for which we predict  $\dot{M_{2}}$ large enough to explain the frequent X-ray outbursts displayed by this source. This limit results in $M_{X}\ga7M_{\odot}$. From the maximum mass solution we find $q\leq0.125$. On the other hand, we have also shown that the stripped-giant model successfully explains the observable properties ($M_2$, $R_2$, $T_{eff}$ and $\dot{M}_2$) of the akin LMXB XTE J1550-564.\\

\section*{Acknowledgments}
JC acknowledges support from the Spanish Ministry of Science and
Technology through the project AYA2006-10254


\bsp

\label{lastpage}

\end{document}